# Liquid-Crystal Transitions: A First Principles Multiscale Approach

Z. Shreif[1], S. Pankavich[1,2], and P. Ortoleva[1]

[1]Center for Cell and Virus Theory
Department of Chemistry
Indiana University
Bloomington IN 47405

[2]Department of Mathematics
University of Texas at Arlington
Arlington, TX 76019 USA

**Abstract**

A rigorous theory of liquid-crystal transitions is developed starting from the Liouville equation. The starting point is an all-atom description and a set of order parameter field variables that are shown to evolve slowly via Newton's equations. The separation of timescales between that of atomic collision/vibrations and the order parameter fields enables the derivation of rigorous equations for stochastic order parameter field dynamics. When the fields provide a measure of the spatial profile of the probability of molecular position, orientation, and internal structure, a theory of liquid-crystal transitions emerges. The theory uses the all-atom/continuum approach developed earlier to obtain a functional generalization of the Smoluchowski equation wherein key atomic details are embedded. The equivalent non-local Langevin equations are derived and the computational aspects are discussed. The theory enables simulations that are much less computationally intensive than molecular dynamics and thus does not require oversimplification of the system's constituent components. The equations obtained do not include factors that require calibration and can thus be applicable to various phase transitions which overcomes the limitations of phenomenological field models. The relation of the theory to phenomenological descriptions of Nematic and Smectic phase transitions, and the possible existence of other types of transitions involving intermolecular structural parameters are discussed.

# I    Introduction

Condensed media composed of phospholipids or other elongated molecules can support liquid crystal phases [1-3]. In these phases, there is preferred molecular orientational order but, unlike crystals, there is no long-lived nearest-neighbor connectivity. The objective of this study is to place these phenomena in the framework of all-atom multiscale statistical mechanics.

Liquid crystals are key elements of many engineered and natural systems. Examples of engineered systems include liquid crystal displays (such as TVs, LCD projectors, and computer monitors), liquid crystal thermometers, and sensors. Lyotropic (concentration-dependent) liquid crystals exist abundantly in biological systems; examples include cell membranes, enveloped viruses, and nanocapsules for therapeutic delivery. It has been suggested that the liquid-crystal properties of these biological systems play a key role in their function [3]. For instance, because of their liquid-crystal properties, membrane phospholipids can easily exchange position while keeping long-lived orientational order on the average. Thus, a predictive theory of the onset and stability of liquid-crystal behavior is of great fundamental and applied interest.

There is an extensive literature on the theory of liquid-crystal phenomena [4-8]. There has been a general consensus that there must be a tradeoff between the level of detail and efficiency in simulating liquid crystals in order to achieve long time and length-scale results. At the highest level of detail, there are the atomistic models [5-6]. These yield insight into nematic and smectic phases of liquid crystals; however, they are unable to simulate long timescales (over 50 ns) or large liquid-crystal systems (such as polymers). Alternatively, different levels of coarse-graining were attempted in order to achieve longer time and length scales. These include molecular simulations using lattice and off-lattice models [7]. These models do not take into consideration the atomistic details and are still not efficient enough to simulate large systems over long time periods (over 1 ms). At the next level of coarse graining are mesoscopic models such as the continuum and Lattice Boltzmann models [8]. In summary, each of the approaches mentioned above has been successful in shedding light on specific aspects of liquid-crystal behavior at the various phases. However, if the aim is to simulate the full behavior of a large system displaying liquid-crystal states among other properties (such as the case of an enveloped virus), then it is necessary to develop a theory that can bridge the gap between the various methods to yield a model that takes into account both atomistic and continuum scales.

Here we present a first-principle theory of liquid-crystals. Starting with Newton's equations for the *N*-atom system, we derive equations for coarse-grained field variables (order parameter fields). The *N*-atom description is necessary to account for key atomic details; however, the liquid-crystal is understood in terms of more coarse-grained variables quantifying the preferred orientation of the elongated constituent molecules. Due to the lack of nearest-neighbor connectivity in such systems, field variables (i.e. variables that change continuously across the system) are needed to keep track of the constantly changing structures. The above prescription is achieved via an integrative all-atom/continuum multiscale (ACM) framework [9-10].

Other methods have been developed wherein an *N*-atom description of the system is used to obtain the coarse-grained model. This is usually done via "decoupled" coarse-graining wherein atoms are lumped together and effective forces on the lumped elements are set forth before simulation begins [11-12]. The shortcoming of such models is that they ignore the feedback loop of Fig.1. Obtaining the forces driving the dynamics of the coarse-grained variables (i.e. the order parameters) only at the beginning of the simulation is insufficient; while the atomic configuration mediates the forces and diffusion coefficients needed to drive the dynamics of the order parameters, the latter affect the ensemble of atomistic configurations. Thus, the former should be co-evolved with the order parameters during a simulation.

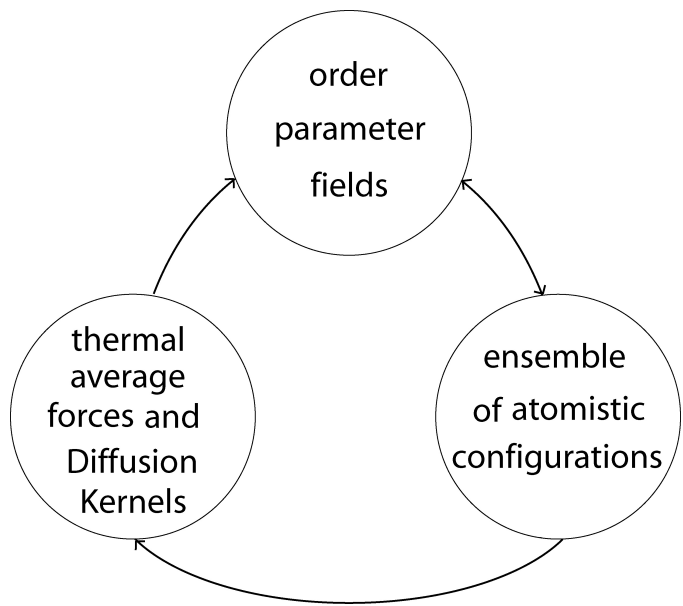

**Fig.1:** Order parameter fields characterizing nanoscale features affect the statistics of the atomistic configurations which, in turn, mediate the forces driving the order parameter fields' dynamics. This feedback loop is central to a complete multiscale understanding of nanosystems and the true nature of their dynamics [13].

Multiscale techniques and their application to Brownian motion have been discussed extensively in the literature [14-24], and date back to the first mathematical explanation of Brownian motion by Einstein and Smoluchowski independently. Most relevant to the work presented here are recent studies [13,25-33] wherein the *N*-atom Liouville equation is used as a starting point to derive reduced equations for the stochastic dynamics of the order parameters. The *N*-atom probability density is written as a function of the all-atom state both directly and, through a set of order parameters, indirectly. This formulation does not constitute an over counting of the number of degrees of freedom $(6N)$; rather, it was shown [13,29-30] that when there is a separation of timescales between the rapidly fluctuating atomic state and the order parameters, both dependencies can be reconstructed. These studies led to our recent development of an ACM (All-atom/Continuum Multiscale) approach [9-10] where the *N*-atom probability density is postulated to be a function of the 6*N* atomic positions and momenta and a functional of a set of order parameter fields. ACM was demonstrated via an application to enveloped viruses [9] and nanocapsule therapeutic delivery [10]. However, that formulation of ACM does not account for liquid-crystals, i.e. does not describe preferred molecular orientation.

In this study, we introduce order parameter field variables that describe the local state of preferred elongated molecule orientational order (Sect II). Starting with the Liouville equation for the *N*-atom probability density $\rho$, we generalize our deductive multiscale procedure [9-10,13] and derive a stochastic equation for the reduced probability density as a functional of the order parameter field profiles (Sect III). In Sect IV, we set forth the equivalent Langevin equations needed for practical simulation of liquid-crystal transitions. In Sect V, we discuss how to proceed to develop a computer simulator for stochastic field dynamics and draw conclusions in Sect VI.

## II      Order Parameter Fields

Counters for an incremental range of molecular structural variables are set forth as a first element of a theory of liquid-crystal transitions as follows. Let $\bar{\Phi}_{jk\underline{\ell}}\left(\Gamma_{jk}^{r}\right)$ be a structural parameter describing the state of the $j^{th}$ molecule of type $k$, where $\Gamma_{jk}^{r}$ is the set of positions of the atoms constituting the $j^{th}$ molecule of type $k$, and $\underline{\ell}$ labels the type of structural parameter (such as center-of-mass position and orientation [13,28]). The notation $\Phi_{jk}$ represents the set of structural parameters with the number of components in the set depending on $k$ (e.g. a large complex flexible molecule requires more structural variables to describe it than does a small rigid one). In other words, each type $k$ has a specific number of distinct $\underline{\ell}$ values, each labeling a distinct type of structural variable. For example, we take $\underline{\ell} = (0,0,0)$ to label the center-of-mass (CM) and $\underline{\ell} = (1,0,0)$ to measure compression/extension along the X-axis. By convention, we take $\underline{0}$ to indicate the CM. With this, $\bar{\Phi}_{jk\underline{0}}$ is the CM of the $j^{th}$ molecule of type $k$.

To obtain the structural parameters, orthogonal basis functions $U_{\underline{\ell}}$ are introduced [9]. Atom $i$ is moved via a deformation wherein its position in space $\vec{r}_i$ is considered a displacement from a reference point $\vec{r}_i^{\,0}$. With this, $\vec{r}_i$ is written

$$\vec{r}_i = \sum_{k=1}^{N_t} \sum_{j=1}^{N_k} \sum_{\underline{\ell}} \bar{\Phi}_{jk\underline{\ell}} U_{\underline{\ell}}\left(\vec{r}_i^{\,0}\right) \lambda_{ijk} + \vec{\sigma}_i, \tag{II.1}$$

where $N_k$ is the number of molecules of type $k$; $N_t$ is the number of molecular types; $\vec{\sigma}_i$ is the residual displacement of atom $i$; and $\lambda_{ijk}$ is one when atom $i$ belongs to the $j^{\text{th}}$ molecule of type $k$ and zero otherwise. The ideal values of the structural parameters are those that minimize the information content (see below), with proper choice of basis functions and number of terms in the $\underline{\ell}$ sum. We start by defining a mass-weighted mean square residual $S$ via

$$S = \sum_{i=1}^{N} m_i \left|\vec{\sigma}_i\right|^2, \tag{II.2}$$

where $N$ is the total number of atoms in the system and $m_i$ is the mass of atom $i$. The relationship between the structural parameters $\bar{\Phi}_{jk\underline{\ell}}$ and the set of atomic positions $\Gamma_r = \{\vec{r}_1, \vec{r}_2, \cdots \vec{r}_N\}$ can be found via minimizing $S$ with respect to $\bar{\Phi}_{jk\underline{\ell}}$ keeping $\Gamma_r$ constant. With this, one obtains

$$\sum_{\underline{\ell}'} \bar{\Phi}_{jk\underline{\ell}'} B_{\underline{\ell}\underline{\ell}'}^{jk} = \sum_{i=1}^{N} m_i \vec{r}_i U_{\underline{\ell}}\left(\vec{r}_i^{\,0}\right) \lambda_{ijk}, \tag{II.3}$$

$$B_{\underline{\ell}\underline{\ell}'}^{jk} = \sum_{i=1}^{N} m_i U_{\underline{\ell}}\left(\vec{r}_i^{\,0}\right) U_{\underline{\ell}'}\left(\vec{r}_i^{\,0}\right) \lambda_{ijk}. \tag{II.4}$$

By convention, we take $U_{\underline{0}} = 1$ and thus $B_{\underline{0}\underline{0}}^{jk}$ is the total mass of a molecule of type $k$. We construct the $U_{\underline{\ell}}$ for $\underline{\ell} \neq \underline{0}$ to be mass-weighted orthogonal to $U_{\underline{0}}$. With this, $B_{\underline{0}\underline{\ell}}^{jk} = B_{\underline{\ell}\underline{0}}^{jk} = 0$ for $\underline{\ell} \neq \underline{0}$. This convention with equation (II.3) implies that $\bar{\Phi}_{jk\underline{0}}$ is the CM of the $j^{\text{th}}$ molecule of type $k$. This procedure was applied in another context (notable for modeling bionanostructures) and it was found that the mass-weighted residual method yields structural parameters similar to generalized CMs [13].

Let $\bar{\varphi}_{k\underline{\ell}}$ be a given value of a structural parameter, i.e. $\bar{\varphi}_{k\underline{\ell}}$ does not depend on the set of atomic positions $\Gamma_r$. The subscript $k$ indicates that the $\underline{\ell}$ value only includes structural variables for a type $k$ molecule. The notation $\varphi_k$ corresponds to the set of $\bar{\varphi}_{k\underline{\ell}}$ values, i.e. for all $\underline{\ell}$ relevant for type $k$ molecules.

Let $\Theta_k$ be a window function centered about zero such that it is one inside the window and zero otherwise. The index $k$ on $\Theta_k$ is introduced because the dimension of the window is the number of structural variables for a molecule of type $k$ which, henceforth, does not include the CM. Therefore, $\Theta_k(\varphi_k - \Phi_{jk}) = 1$ indicates that the $j^{th}$ molecule of type $k$ has internal structural variables $\Phi_{jk}$ in a window about $\varphi_k$.

Order parameter fields are conceived of as densities of subpopulations of molecules of type $k$ with structural variables in a window. If a variable has a narrow window, then fluctuations of molecules in and out of the window will be large and occur on a short timescale; therefore the associated subpopulation density is not a viable order parameter. Let the length scale over which the profile of the CM distribution be broad, i.e. we seek a smoothly varying spatial profile of subpopulation CM distribution. Thus, we introduce a set of order parameter fields $\Upsilon_k$ for type $k$ molecules such that

$$\Upsilon_k(\Gamma_r, \vec{R}, \varphi_k) = \sum_{j=1}^{N_k} \delta(\vec{R} - \varepsilon \vec{\Phi}_{jk\underline{0}}) \Theta_k(\varphi_k - \Phi_{jk}), \qquad (II.5)$$

where $\delta$ is the Dirac delta function centered at $\vec{0}$ (which, for formal development, we take to be a narrow unit-normalized, gaussian); the parameter $\varepsilon$ is the ratio of the average nearest-neighbor molecule CM distance to the characteristic size of features of interest; $\vec{R}$ labels a point in space and is scaled such that it undergoes a displacement of $\varepsilon$ when one average nearest-neighbor distance is traversed; and $\Upsilon_k(\Gamma_r, \vec{R}, \varphi_k)$ times the mass of a type $k$ molecule is the mass density of type $k$ molecules with internal structure in the $\Theta_k$ window and CM near $\varepsilon^{-1}\vec{R}$.

The first step in our multiscale development is to show that the densities $\Upsilon_k$ are slowly varying in time and thus, can be used as order parameters [13]. Newton's equations imply $d\Upsilon_k/dt = -\mathcal{L}\Upsilon_k$ for Liouville operator $\mathcal{L}$

$$\mathcal{L} = -\sum_{i=1}^{N} \left\{ \frac{\vec{p}_i}{m_i} \cdot \frac{\partial}{\partial \vec{r}_i} + \vec{F}_i \cdot \frac{\partial}{\partial \vec{p}_i} \right\} \qquad (II.6)$$

where $\vec{p}_i$ is the momentum of atom $i$ and $\vec{F}_i$ is the force exerted on it. This yields

$$d\Upsilon_k/dt = -\sum_{j=1}^{N_k} \sum_{i=1}^{N} \frac{\vec{p}_i}{m_i} \cdot \vec{\pi}_{ijk} \qquad (II.7)$$

$$\bar{\pi}_{ijk} = \varepsilon \frac{\partial \bar{\Phi}_{jk\underline{0}}}{\partial \bar{r}_i} \frac{\partial \delta(\bar{R} - \varepsilon \bar{\Phi}_{jk\underline{0}})}{\partial \bar{R}} \Theta_k(\varphi_k - \Phi_{jk}) + \sum_{\underline{\ell}} \frac{\partial \bar{\Phi}_{jk\underline{\ell}}}{\partial \bar{r}_i} \delta(\bar{R} - \varepsilon \bar{\Phi}_{jk\underline{\ell}}) \frac{\partial \Theta_k(\varphi_k - \Phi_{jk})}{\partial \varphi_{k\underline{\ell}}}. \tag{II.8}$$

Multiplying both sides of (II.7) by a sampling function $f(\bar{R} - \tilde{\bar{R}})$, for a given point in space $\tilde{\bar{R}}$, and integrating over a volume in $\bar{R}$-space, we find that $\Upsilon_k$ is slowly varying due to the large number of molecules in such a volume; by examining the LHS of (II.7), one obtains

$$\text{LHS} = \frac{\mathrm{d}}{\mathrm{d}t} \int \mathrm{d}^3 R \, \Upsilon_k(\Gamma_r, \bar{R}, \varphi_k) f(\bar{R} - \tilde{\bar{R}}) \tag{II.9}$$

which is of the order of the number of molecules of type $k$ in a given structural window in the sampling volume. If the sampling volume is large enough, the number of molecules would also be large. Choosing the sampling volume such that the typical number of molecules of a given type in a structural window is $\mathrm{O}(\varepsilon^{-2})$, then (II.9) is of $\mathrm{O}(\varepsilon^{-2})$. By examining the RHS of (II.7), and using integration by parts, one obtains

$$\text{RHS} = -\sum_{j=1}^{N_k} \sum_{i=1}^{N} \frac{\bar{p}_i}{m_i} \cdot \int \mathrm{d}^3 R \, \bar{\pi}_{ijk} f(\bar{R} - \tilde{\bar{R}}) \tag{II.10}$$

where

$$\int \mathrm{d}^3 R \, \bar{\pi}_{ijk} f(\bar{R} - \tilde{\bar{R}}) = \varepsilon \frac{\partial \bar{\Phi}_{jk\underline{0}}}{\partial \bar{r}_i} \Theta_k \frac{\partial}{\partial \tilde{\bar{R}}} \int \mathrm{d}^3 R \, \delta(\bar{R} - \varepsilon \bar{\Phi}_{jk\underline{0}}) f(\bar{R} - \tilde{\bar{R}}) \tag{II.11}$$

$$+ \sum_{\underline{\ell}} \frac{\partial}{\partial \varphi_{k\underline{\ell}}} \frac{\partial \bar{\Phi}_{jk\underline{\ell}}}{\partial \bar{r}_i} \Theta_k \int \mathrm{d}^3 R \, \delta(\bar{R} - \varepsilon \bar{\Phi}_{jk\underline{0}}) f(\bar{R} - \tilde{\bar{R}})$$

The integral in the RHS of (II.11) is of the order of the number of molecules in the sampling volume, that is $\mathrm{O}(\varepsilon^{-2})$. However, (II.10) is a sum of terms the number of which is on the order of the number of molecules. Therefore, (II.10) is a sum of fluctuating terms of $\mathrm{O}(\varepsilon^{-2})$ each of order $\mathrm{O}(\varepsilon^{-2})$; this makes the overall order of (II.10) of $\mathrm{O}(\varepsilon^{-1})$. This suggests introducing a scaled order parameter $\psi_k = \varepsilon \Upsilon_k$. With this, we obtain

$$\mathrm{d}\psi_k / \mathrm{d}t = -\varepsilon \sum_{j=1}^{N_k} \sum_{i=1}^{N} \frac{\bar{p}_i}{m_i} \cdot \bar{\pi}_{ijk} \equiv \varepsilon \Pi_k(\Gamma, \bar{R}, \varphi_k), \tag{II.12}$$

where $\Gamma = \{\Gamma_r, \vec{p}_1, \vec{p}_2, \cdots \vec{p}_N\}$ is the set of $6N$ atomic positions and momenta. The field variables $\psi_k$ and $\Pi_k$ are labeled by $\vec{R}$ and $\varphi_k$ and vary continuously across the system. Thus, with the dependency on the atomistic configurations being understood, we use the notation $\psi_k(\vec{R}, \varphi_k)$ and $\Pi_k(\vec{R}, \varphi_k)$ to label them as the $(\vec{R}, \varphi_k)$-associated parameters. In the following sections, we use the notation $\Psi_k$ to represent the infinite set composed of $\psi_k(\vec{R}, \varphi_k)$ for all values of $\vec{R}$ and $\varphi_k$.

## III    Multiscale Analysis: A Functional Smoluchowski Equation for Liquid Crystals

To develop a theory for liquid crystals using the field variables introduced in the previous section, we derive an equation for the reduced probability density $W$ that is a functional of the distribution of field variables in space $\vec{R}$ and over the continuous set of structural parameters $\varphi_k$. To address the continuum limit of the field theory we present, we divide the system into a large but finite number of boxes of volume $v_c$ (each of which is labeled by its center point $\vec{R}$). As indicated previously [10], $v_c$ is taken to be much smaller than the characteristic length of the phenomena of interest, but large enough to contain a statistically significant number of molecules. By definition,

$$W[\underline{\Psi}, t] = \int d\Gamma^* \Delta(\underline{\Psi} - \underline{\Psi}^*) \rho(\Gamma^*, t), \tag{III.1}$$

where

$$\Delta(\underline{\Psi} - \underline{\Psi}^*) = \prod_{k=1}^{N_t} \Delta_k(\Psi_k - \Psi_k^*); \tag{III.2}$$

$\Gamma^*$ is the $N$-atom state over which integration is taken; $\underline{\Psi} = \{\Psi_1, \Psi_2, \cdots \Psi_{N_t}\}$; a superscript * for any variable indicates evaluation at $\Gamma^*$; $\rho$ is the $N$-atom probability density which satisfies the Liouville equation $\partial \rho / \partial t = \mathcal{L} \rho$; and $\Delta_k$ is a continuum product of $\delta(\psi_k - \psi_k^*)$ for all positions $\vec{R}$ and set of structural parameters $\varphi_k$, and is represented by

$$\Delta_k = \exp\left\{\frac{1}{v_c}\int d^3R d\varphi_k \ln\left[\delta\left(\psi_k(\vec{R},\varphi_k) - \psi_k^*(\vec{R},\varphi_k)\right)\right]\right\}.$$ With these definitions and the chain rule for functionals, $W$ is found to satisfy the conservation equation

$$\frac{\partial W}{\partial t} = -\frac{\varepsilon}{v_c}\sum_{k=1}^{N_I}\int d^3R d\varphi_k \frac{\delta}{\delta\psi_k(\vec{R},\varphi_k)}\int d\Gamma^* \Delta(\underline{\Psi}-\underline{\Psi}^*)\Pi_k^*(\vec{R},\varphi_k)\rho(\Gamma^*,t), \qquad (\text{III}.3)$$

Given the ansatz that $\rho$ depends on $\Gamma$ both directly and, through the order parameter fields, indirectly, $\rho$ is rewritten as a functional of $\underline{\Psi}$ and a function of $\Gamma$, i.e.

$$\rho \equiv \rho[\Gamma,\underline{\Psi},t_0,\underline{t};\varepsilon]. \qquad (\text{III}.4)$$

The set of scaled times $t_n = \varepsilon^n t$, $n = 0,1,\cdots$ is introduced to account for processes taking place on increasingly long timescales; and $\underline{t} = \{t_1, t_2, \cdots\}$ is the set of times tracking the slow processes, i.e. much slower than the atomic processes tracked by $t_0$. With this and the chain rule, we obtain the multiscale Liouville equation

$$\sum_{n=0}^{\infty}\varepsilon^n\frac{\partial\rho}{\partial t_n} = (\mathcal{L}_0 + \varepsilon\mathcal{L}_1)\rho \qquad (\text{III}.5)$$

$$\mathcal{L}_0 = -\sum_{i=1}^{N}\left(\frac{\vec{p}_i}{m_i}\cdot\frac{\partial}{\partial\vec{r}_i} + \vec{F}_i\cdot\frac{\partial}{\partial\vec{p}_i}\right) \qquad (\text{III}.6)$$

$$\mathcal{L}_1 = -\frac{1}{v_c}\sum_{k=1}^{N_I}\int d^3R d\varphi_k \Pi_k(\vec{R},\varphi_k)\frac{\delta}{\delta\psi_k(\vec{R},\varphi_k)}. \qquad (\text{III}.7)$$

In $\mathcal{L}_0$ the $\Gamma$-derivatives are taken at constant $\underline{\Psi}$, while in $\mathcal{L}_1$ the functional derivatives are taken at constant $\Gamma$. Following our earlier procedure [9-10], our strategy is to solve the Liouville equation perturbatively in the small $\varepsilon$ limit, and then use the solution to obtain a closed equation for the stochastic dynamics of the field variables. The development continues with the perturbation expansion

$$\rho = \sum_{n=0}^{\infty}\varepsilon^n\rho_n, \qquad (\text{III}.8)$$

and examining the multiscale Liouville equation at each order in $\varepsilon$. We hypothesize that the lowest order solution $\rho_0$ is slowly varying in time and is, thus, independent of $t_0$.

To $O(\varepsilon^0)$, the above assumptions imply $\mathcal{L}_0 \rho_0 = 0$. Since no further information is known, $\rho_0$ is determined by adopting an entropy maximization procedure with canonical constraint of fixed average energy [13]. With this, we obtain

$$\rho_0[\Gamma;\underline{\Psi},\underline{t}] = \hat{\rho}[\Gamma;\underline{\Psi}]W_0[\underline{\Psi},\underline{t}] \tag{III.9}$$

$$\hat{\rho} = \frac{\exp(-\beta H)}{Q[\underline{\Psi}]} \tag{III.10}$$

where $\beta$ is inverse temperature, $H$ is the Hamiltonian

$$H(\Gamma) = \sum_{i=1}^{N} \frac{p_i^2}{2m_i} + V(\Gamma_r) \tag{III.11}$$

for $N$-atom potential $V$, and $Q$ is the $\underline{\Psi}$-dependent partition functional given by

$$Q[\underline{\Psi}] = \int d\Gamma^* \Delta(\underline{\Psi} - \underline{\Psi}^*) \exp(-\beta H^*). \tag{III.12}$$

To $O(\varepsilon)$, the multiscale Liouville equation implies

$$\left(\frac{\partial}{\partial t_0} - \mathcal{L}_0\right)\rho_1 = -\frac{\partial \rho_0}{\partial t_1} + \mathcal{L}_1 \rho_0. \tag{III.13}$$

This admits the solution

$$\rho_1 = -\int_0^{t_0} dt_0' e^{\mathcal{L}_0(t_0 - t_0')} \left\{\frac{\partial \rho_0}{\partial t_1} - \mathcal{L}_1 \rho_0\right\}, \tag{III.14}$$

where the initial first order distribution was taken to be zero (i.e. the system is in a quasi-equilibrium state at the initial instant) [29-30].

Inserting (III.7) and (III.9) in (III.14), we obtain

$$\rho_1 = -t_0 \hat{\rho} \frac{\partial W_0}{\partial t_1} - \hat{\rho} \int_0^{t_0} dt_0' e^{\mathcal{L}_0(t_0 - t_0')} \sum_{k=1}^{N_r} \frac{1}{v_c} \int d^3 R d\varphi_k \, \Pi_k(\vec{R},\varphi_k) \left(\frac{\delta}{\delta \psi_k(\vec{R},\varphi_k)} - \beta \langle h_k(\vec{R},\varphi_k)\rangle\right) W_0, \tag{III.15}$$

where the thermal-average $\langle h_k \rangle$ is defined via

$$\langle h_k(\vec{R},\varphi_k)\rangle = -\frac{\delta F[\underline{\Psi}]}{\delta \psi_k(\vec{R},\varphi_k)}, \tag{III.16}$$

and $F$ is the free energy, related to $Q$ via

$$Q = e^{-\beta F}. \tag{III.17}$$

Using the Gibbs hypothesis, imposing that $\rho_1$ be finite as $t_0 \to \infty$, and using the fact that the thermal-average of $\Pi_k$ is zero (since the weighing factor $\hat{\rho}$ is even in $\vec{p}_i$ while $\Pi_k$ is odd in it), we conclude that $W_0$ is independent of $t_1$. With this, (III.15) becomes

$$\rho_1 = -\hat{\rho}\int_0^{t_0} dt_0' e^{\mathcal{L}_0(t_0-t_0')} \left\{ \sum_{k=1}^{N_t} \frac{1}{v_c} \int d^3R d\varphi_k \; \Pi_k(\vec{R},\varphi_k) \left( \frac{\delta}{\delta\psi_k(\vec{R},\varphi_k)} - \beta\langle h_k(\vec{R},\varphi_k)\rangle \right) \right\} W_0. \quad \text{(III.18)}$$

Inserting (III.9) and (III.18) in the conservation equation (III.3) and using the fact that $W \to W_0$ as $\varepsilon \to 0$ yields

$$\frac{\partial W}{\partial t} = \varepsilon^2 \sum_{k,k'=1}^{N_t} \int \frac{d^3R}{v_c} \frac{d^3R'}{v_c} d\varphi_k d\varphi_{k'} \frac{\delta}{\delta\psi_k(\vec{R},\varphi_k)} \left\{ D_{kk'} \left( \frac{\delta}{\delta\psi_{k'}(\vec{R}',\varphi_{k'})} - \beta\langle h_{k'}(\vec{R}',\varphi_{k'})\rangle \right) \right\} W, \quad \text{(III.19)}$$

where the diffusion coefficients $D_{kk'}$ are defined as

$$D_{kk'}(\vec{R},\vec{R}',\varphi_k,\varphi_{k'}) = \int_{-\infty}^{0} dt_0 \langle \Pi_k(\vec{R},\varphi_k) e^{-\mathcal{L}_0 t_0} \Pi_{k'}(\vec{R},\varphi_{k'})\rangle. \quad \text{(III.20)}$$

This constitutes a Smoluchowski equation for the stochastic dynamics of the field variables and can thus provide a theory of liquid-crystal phenomena. The set of Langevin equations equivalent to (III.19) allow a practical simulation as outlined in the next section.

## IV  Langevin Equations of Liquid-Crystal Transition

For practical simulation of Liquid-Crystal transition, we derive the Langevin equations equivalent to the Smoluchowski equation (III.19). We suggest the Langevin equations take the form

$$\frac{\partial \psi_k(\vec{R},\varphi_k)}{\partial t_2} = \frac{1}{v_c} \sum_{k'=1}^{N_t} \int d^3R' d\varphi_{k'} \gamma_{kk'}(\vec{R},\vec{R}',\varphi_k,\varphi_{k'}) \langle h_{k'}(\vec{R}',\varphi_{k'})\rangle + \xi_k(t_2,\vec{R},\varphi_k) \quad \text{(IV.1)}$$

where $\xi_k$ are random forces and $\gamma_{kk'}$ are kernels which we will show to be related to $D_{kk'}$ of Sect. III. In what follows, we demonstrate that this implies the Smoluchowski equation (III.19). For simplicity, we henceforth drop the subscript 2 on $t_2$.

Starting with the Markov assumption and assuming the set of random forces $\underline{\xi} = \{\xi_1, \xi_2, \cdots \xi_{N_t}\}$ changes much more rapidly than the set of field order parameters $\underline{\Psi}$, $W$ is postulated to evolve via

$$W[\underline{\Psi}, t+\tau] = \mathcal{S}_{\underline{\Psi}'} \mathcal{S}_{\underline{\xi}} T[\underline{\xi}, t, \tau] \Delta(\underline{\Psi} - \underline{\tilde{\Psi}}) W[\underline{\Psi}', t] \qquad (IV.2)$$

where $\mathcal{S}$ is a functional integral, $T[\underline{\xi}, t, \tau]$ is the transition probability functional for a scenario of $\underline{\xi}$ during the time interval $t$ to $t+\tau$; $\underline{\tilde{\Psi}} = \{\tilde{\Psi}_1, \tilde{\Psi}_2, \cdots \tilde{\Psi}_{N_t}\}$; $\underline{\Psi}' = \{\Psi'_1, \Psi'_2, \cdots \Psi'_{N_t}\}$; $\Psi'_k$ is the set of $\psi'_k$ for all $\vec{R}$ and $\varphi_k$; $\tilde{\Psi}_k$ is the set of $\tilde{\psi}_k$ for all $\vec{R}$ and $\varphi_k$; and $\tilde{\psi}_k(\vec{R}, \varphi_k)$ is the solution to (IV.1) at time $t+\tau$ given that the state was $\psi'_k$ at time $t$.

Assuming that $W$ changes very little over $\tau$, we get

$$W[\underline{\Psi}, t+\tau] = W[\underline{\Psi}, t] + \frac{\partial W}{\partial t}\tau + \cdots \qquad (IV.3)$$

Inserting (IV.3) in the LHS of (IV.2) and expanding the $\Delta$-function on the RHS in $(\underline{\Psi} - \underline{\Psi}')$ truncated at second order (following similar procedure as earlier [10]), we obtain

$$\frac{\partial W}{\partial t}\tau + \cdots = \sum_{k=1}^{N_t} \int \frac{d^3 R}{v_c} d\varphi_k \frac{\delta}{\delta \psi_k(\vec{R}, \varphi_k)} \langle \psi_k(\vec{R}, \varphi_k) - \tilde{\psi}_k(\vec{R}, \varphi_k) \rangle_T W \qquad (IV.4)$$

$$+ \frac{1}{2} \sum_{k,k'=1}^{N_t} \int \frac{d^3 R}{v_c} \frac{d^3 R'}{v_c} d\varphi_k d\varphi_{k'} \left\{ \frac{\delta^2}{\delta \psi_k(\vec{R}, \varphi_k) \delta \psi_{k'}(\vec{R}', \varphi_{k'})} \right.$$

$$\left. \langle (\psi_k(\vec{R}, \varphi_k) - \tilde{\psi}_k(\vec{R}, \varphi_k))(\psi_{k'}(\vec{R}', \varphi_{k'}) - \tilde{\psi}_{k'}(\vec{R}', \varphi_{k'})) \rangle_T \right\} W$$

where $\langle A \rangle_T \equiv \mathcal{S}_{\underline{\xi}} T[\underline{\xi}, t, \tau] A$ is the $T$-weighted average for any $\underline{\xi}$-dependent variable $A$. In what follows, we assume that $\langle \xi_k(t, \vec{R}, \varphi_k) \rangle_T = 0$ and that the process generating $\xi_k$ is stationary. With this, integrating (IV.1) and neglecting terms of order $\tau^2$ and higher yields

$$\langle \psi_k(\vec{R}, \varphi_k) - \tilde{\psi}_k(\vec{R}, \varphi_k) \rangle_T = -\tau \sum_{k'=1}^{N_t} \frac{1}{v_c} \int d^3 R d\varphi_k \gamma_{kk'}(\vec{R}, \vec{R}', \varphi_k, \varphi_{k'}) \langle h_{k'}(\vec{R}', \varphi_{k'}) \rangle \qquad (IV.5)$$

$$\langle (\psi_k(\vec{R}, \varphi_k) - \tilde{\psi}_k(\vec{R}, \varphi_k))(\psi_{k'}(\vec{R}', \varphi_{k'}) - \tilde{\psi}_{k'}(\vec{R}', \varphi_{k'})) \rangle_T = \tau J_{kk'}(\vec{R}, \vec{R}', \varphi_k, \varphi_{k'}) \qquad (IV.6)$$

where

$$J_{kk'}(\vec{R},\vec{R}',\varphi_k,\varphi_{k'}) = \int_{-\infty}^{0} ds\, \phi_{kk'}(s,\vec{R},\vec{R}',\varphi_k,\varphi_{k'}) \tag{IV.7}$$

$$\phi_{kk'}(t-t',\vec{R},\vec{R}',\varphi_k,\varphi_{k'}) = \left\langle \xi_k(t,\vec{R},\varphi_k)\xi_{k'}(t',\vec{R}',\varphi_{k'}) \right\rangle_T. \tag{IV.8}$$

Inserting (IV.5) and (IV.6) in (IV.4) yields

$$\frac{\partial W}{\partial t} = \sum_{k,k'=1}^{N_r} \int \frac{d^3 R}{v_c} \frac{d^3 R'}{v_c} d\varphi_k d\varphi_{k'} \frac{\delta}{\delta \psi_k(\vec{R},\varphi_k)} \left\{ \frac{J_{kk'}}{2} \frac{\delta}{\delta \psi_{k'}(\vec{R}',\varphi_{k'})} - \gamma_{kk'} \left\langle h_{k'}(\vec{R}',\varphi_{k'}) \right\rangle \right\} W. \tag{IV.9}$$

Comparing (IV.9) with the functional Smoluchowski equation (III.19), we find that the $\gamma_{kk'}$ kernels, the diffusion coefficients $D_{kk'}$, and the statistics of the random forces are related via

$$\gamma_{kk'}(\vec{R},\vec{R}',\varphi_k,\varphi_{k'}) = \beta D_{kk'}(\vec{R},\vec{R}',\varphi_k,\varphi_{k'}) \tag{IV.10}$$

$$D_{kk'}(\vec{R},\vec{R}',\varphi_k,\varphi_{k'}) = \frac{1}{2}\int_{-\infty}^{0} ds \left\langle \xi_k(0,\vec{R},\varphi_k)\xi_{k'}(s,\vec{R}',\varphi_{k'}) \right\rangle. \tag{IV.11}$$

The Langevin equations presented in this section provide a practical way to simulate stochastic liquid crystal dynamics wherein atomic details are retained via the use of the diffusion coefficients and the thermal average forces. The latter are to be numerically calculated via molecular dynamics (MD) simulations and an efficient algorithm for developing the ensembles. In addition, the statistics of the random forces used in the Langevin equations are restricted via (IV.11).

## V     Simulating Stochastic Field Dynamics

The all-atom/continuum multiscale theory holds great promise for the efficient simulation of liquid-crystal transitions. Providing simulations that use the equations derived here is beyond the scope of this study. However, to illustrate the feasibility of a multiscale computational approach, we suggest the following algorithm.

As the set order parameters are $(\vec{R},\varphi_k)$-dependent field variables, the first step is to divide the space of $\vec{R}$ and that of each $\bar{\varphi}_{k\underline{\ell}}$ into finite elements. The order parameter fields $\psi_k(\vec{R},\varphi_k)$ are then defined in each finite volume. The space of $\vec{R}$ is discretized into a cubic grid while each $\bar{\varphi}_{k\underline{\ell}}$ is discretized in a way that depends on $\underline{\ell}$. The size of each element/structural window should be smaller than a critical characteristic size but large enough to contain a statistically significant number of molecules. For example, consider a system consisting of two types of molecules. For the first type, only the center of mass position is of interest. For the second type, information on both center of mass position and orientation is needed. In this case, $\varphi_1 = \{\bar{\varphi}_{10}\}$ and $\varphi_2 = \{\bar{\varphi}_{20}, \bar{\varphi}_{2\ell_1}\}$. The space of $\bar{\varphi}_{2\ell_1}$ is then divided into cones and a domain number $b_1$ is assigned to each cone while a domain number $a$ is assigned to each cube in the space of $\bar{R}$. With this, $\psi_1(a)$ corresponds to the number of type 1 molecules in the cube $a$, while $\psi_2(a,b_1)$ corresponds to the number of type 2 molecules in cube $a$ with an orientation in the cone $b_1$. In general, $\psi_k(a,\underline{b}^k)$ corresponds to the number of type $k$ molecules in the $a^{th}$ cube with characteristics in the set of domains $\underline{b}^k = \{b_i, b_j, \cdots\}$ (i.e. in a structural window defined by the set of domains $\underline{b}^k$). For example, while $b_1$ corresponds to an orientational window, $b_2$ can correspond to overall size, $b_3$ to degree of coiling, etc, and which of these windows is appropriate will depend on the molecular type $k$. Note that the size of each cube $a$ should be large enough to contain several molecules as $\bar{R}$ represents a coarse-grained location and is not intended to capture small scale structure.

The next step is to calculate the thermal-average forces and diffusion coefficients. The thermal-average forces are computed via a method for generating an ensemble of atomistic configurations, followed by a Monte Carlo integration to obtain the thermal-average of the order parameter force calculated from each configuration. The friction factors are computed using MD-generated time courses used to compute correlation functions for the rates of change of the order parameters (see III.20). Even when using a highly efficient parallelized algorithm, the calculations might not be tractable unless the kernels $D_{kk'}(\bar{R},\bar{R}',\varphi_k,\varphi_{k'})$ are short range; in this case, only interactions between nearest-neighbor elements are needed. The thermal-average forces and diffusion coefficients are then used to drive the dynamics of the order parameters. $\psi_k(a,\underline{b}^k)$ at time $t+\tau$ for timestep $\tau$ will then be computed using

$$\psi_k(a,\underline{b}^k,t+\tau) = \psi_k(a,\underline{b}^k,t) + \beta\tau\sum_{k'=1}^{N_t}\sum_{a',\underline{b}^{k'}} D_{kk'}(a,a',\underline{b}^k,\underline{b}^{k'})h_{k'}(a',\underline{b}^{k'}) + \xi_k(a,\underline{b}^k,t) \qquad (V.1)$$

where $\xi_k$ is, for example, a sequence of Gaussian time pulses with random spacing, width, and time where the random numbers are chosen according to probability distributions consistent with the MD-correlation functions. The thermal-average forces and diffusion coefficients are then updated and the cycle is repeated (see Fig. 2).

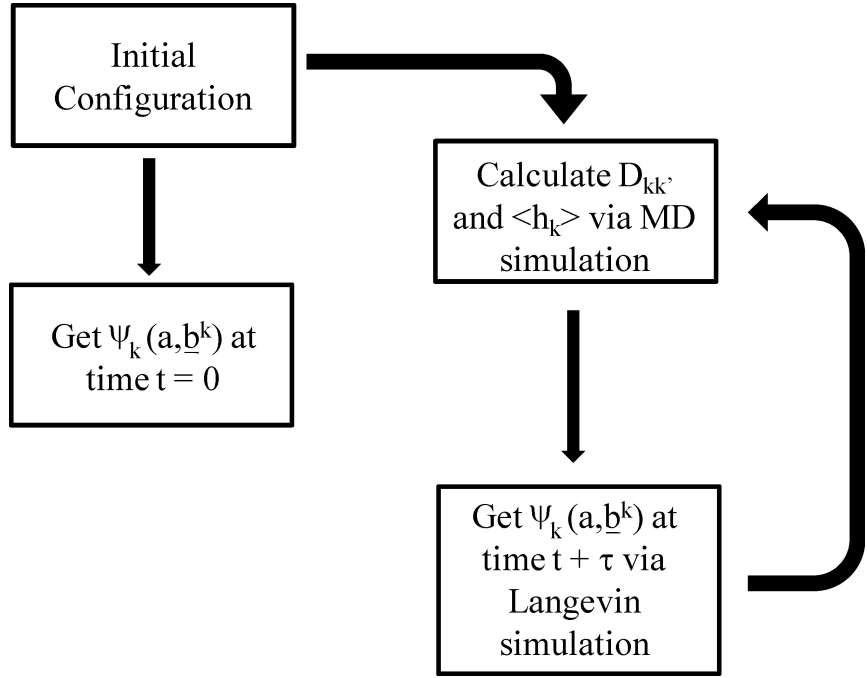

**Fig.2**    An efficient algorithm involves starting with an initial configuration to obtain the initial values of the thermal average forces $\langle h_k \rangle$ and diffusion coefficients $D_{kk'}$, in addition to the structural parameters, and thus the initial values of the order parameter fields. The former are used to drive the dynamics of the order parameter fields. The thermal average forces and diffusion coefficients are then updated "on the fly" and the cycle is repeated. This scheme is consistent

## VI    Conclusions

An all-atom/continuum multiscale (ACM) approach for simulating liquid-crystal transitions is presented. Structural parameters describing the state of each type of molecule are introduced (e.g. center-of-mass, orientation, and compression/extension) and order parameter fields are defined as the densities of subpopulations of molecules whose structural variables reside in a specified window. A functional Smoluchowski equation for the stochastic dynamics of the order parameter fields and its equivalent Langevin equations are derived. An algorithm for simulating these Langevin equations is provided.

Our approach allows for simulation of liquid-crystal transitions that retains key atomic detail and does not need calibration, in contrast to phenomenological models. The theory is built on our ACM [9-10] approach which is tailored to systems without long-lasting nearest-neighbor molecular connectivity but which can support long-lasting organization on the average. ACM was originally developed to investigate the behavior of enveloped viruses [9] and nanocapsules for the delivery of therapeutic agents [10]. However, the liquid-crystal behavior of these two systems was not taken into consideration. With the present approach, a more complete study of both systems can now be conducted.

Computational approaches for liquid crystal transitions have been based on either semi-phenomenological molecular [34] or phenomenological field [4,35-36] models. The former describe the interactions at the molecular levels; however, they simplify the system to make the model computationally feasible (e.g. rigid rod or bead models). Field models, on the other hand, describe the system via order parameter profiles and use a phenomenological expression for the free-energy functional. Take, for example, the widely used de Gennes models [4]. Two different models were introduced, one for the isotropic-nematic transitions and another for the nematic-smecticA one; the reason for having distinct models is that the model that works for one doesn't work for the other. Thus, there was a need for an approach that subsumes multiple types of transitions and possible more complex ones. Phenomenological models that extend the Landau-de Gennes free energy functional to simultaneously describe more than one transition were presented [35-36]. As the free energy expression becomes more complex, more parameters are introduced that need calibration with each new application. This implies that it is possible to introduce a unified phenomenological theory describing the various known transitions; however, such a model does not allow the discovery of new phase transitions as a first principle theory would. Also, the increased complexity and number of calibrated parameters would prove to be restrictive for a computational model and one could question the true predictive character of such a theory.

To illustrate our approach, consider a system that exhibits 3 phases: isotropic-nematic-smecticA. Keeping in mind that one can include as many structural parameters as needed in our theory, at least four are required for this case: center-of-mass $\bar{\Phi}_{jk\underline{0}}$ and orientations $\Phi_{jk\underline{\ell}}$ for 3 $\underline{\ell}$ values, for $j = 1, \cdots N_k$ and for each type $k$ requiring these parameters. Let

$$T_\Omega(\Gamma_r, \varphi_k) \equiv \frac{\int_\Omega d^3R d\varphi'_k \psi_k(\bar{R}, \varphi'_k) \hat{f}(\varphi_k - \varphi'_k)}{\int_\Omega d^3R d\varphi'_k \psi_k(\bar{R}, \varphi'_k)}$$

be the probability that a type $k$ molecule in a sampling volume $\Omega$ has orientations $\varphi_k$ specified by the sampling function $\hat{f}$ (see appendix for the special case of rigid axisymmetric molecules). When the system is in the isotropic liquid phase, $T_\Omega$ is a constant. As the system transitions to the nematic phase, $T_\Omega$ becomes non-constant and indicates preferred orientation. For the transition from the nematic to the smectic phases, one needs to examine the atomistic probability density to extract the short-scale order which is computed via correlation functions using the conditional probability $\hat{\rho}$ (of III.10) and Monte Carlo integration. With the inclusion of other internal molecular structural parameters, new types of transitions can be discovered. For example, the constituent molecules might have a tendency to assume a specific structure (e.g a helix) versus another one.

Our approach has the advantages of the molecular models in the sense that key small scale information is taken into consideration and can thus be applied to multiple types of phase transitions without the above cited difficulties of these models (i.e. without the need to simplify the components composing the system and for recalibration with each new application).

**Appendix**

Consider the case of a single type of molecule so that the subscript $k$ can be dropped for all parameters. If the molecules are rigid, then the residuals $\bar{\sigma}_i$ are zero and the position of atom $i$ is given by

$$\bar{r}_i = \bar{\Phi}_{j(i)\underline{0}} + \sum_{\underline{\ell}}^{rot} \bar{\Phi}_{j(i)\underline{\ell}} U_{\underline{\ell}}(\bar{s}_i^0), \tag{A.1}$$

where $\vec{s}_i = \vec{r}_i - \vec{\Phi}_{j(i)0}$ (i.e. we consider the displacements from the CM of $j(i)$ as they are more convenient), $\vec{s}_i^0$ is a reference point, $\vec{s}_i$ is the relative position after rotation from the reference configuration, and the superscript "rot" on the sum indicates that the three essential basis functions are $U_{100} = s_{ix}^0$, $U_{010} = s_{iy}^0$, and $U_{001} = s_{iz}^0$ where $s_{ix}^0$, $s_{iy}^0$, $s_{iz}^0$ are the components of $\vec{s}_i^0$ along the $x$, $y$, and $z$-axes, respectively. With this, we obtain

$$\vec{s}_i = \vec{\Phi}_{j(i)100} s_{ix}^0 + \vec{\Phi}_{j(i)010} s_{iy}^0 + \vec{\Phi}_{j(i)001} s_{iz}^0. \tag{A.2}$$

Thus, the set of $\vec{\Phi}_{j(i)\underline{\ell}}$ constitutes a rotation matrix $\vec{\vec{X}}_{j(i)}$ such that

$$\vec{s}_i = \vec{\vec{X}}_{j(i)} \vec{s}_i^0 \tag{A.3}$$

$$\vec{\vec{X}} = \begin{pmatrix} \Phi_{x100} & \Phi_{x010} & \Phi_{x001} \\ \Phi_{y100} & \Phi_{y010} & \Phi_{y001} \\ \Phi_{z100} & \Phi_{z010} & \Phi_{z001} \end{pmatrix}. \tag{A.4}$$

Integrating $\Psi$ with respect to $\vec{R}$ over a volume $\Omega$ yields

$$n_\Omega(\varphi) = \sum_{j \text{ in } \Omega} \Theta(\varphi - \Phi_j) \tag{A.5}$$

where $n_\Omega$ is the number of molecules in $\Omega$ with internal structural variables in a window about $\varphi$. $\varphi$ and $\Phi_j$ are the set of elements in $\vec{\vec{X}}$ for any molecule and molecule $j$, respectively.

For axisymmetric molecules, $\Phi_j$ can be related to the set $\omega$ of two polar angles. Letting $d\omega$ be the two-dimensional polar angle volume element, the orientational probability distribution $T_\Omega$ takes the form

$$T_\Omega(\Gamma_r, \omega) = \frac{\int d\omega' n_\Omega(\omega') \hat{f}(\omega - \omega')}{\int d\omega' n_\Omega(\omega')}. \tag{A.6}$$

For isotropic states, $T_\Omega$ is independent of $\omega$, simply being a measure of the size of the window. In a liquid crystal, it takes structure, i.e., it indicates preferred orientation.

To relate our approach to existing models, consider, for example, the Maier-Saupe model for axisymmetric molecules. The proposed Smoluchowski equations [37-38] to study the dynamics of such systems describe the evolution of the orientational probability distribution. Here, we derive a Langevin equation describing the evolution of $T_\Omega$. If exchange of molecules between $\Omega$ and its surroundings are ignored, then we may choose $n_\Omega$ to be normalized (i.e. $\int d\omega\, n_\Omega(\omega) = 1$); in this case, one obtains

$$\frac{\partial T_\Omega(\Gamma_r, \tilde{\omega})}{\partial t} = \int d\omega \frac{\partial n_\Omega(\omega)}{\partial t} \hat{f}(\tilde{\omega} - \omega). \tag{A.7}$$

Assuming spatial symmetry such that the thermal average force is independent of $\bar{R}$ and inserting the Langevin equation of Sect. IV in (A.7) yields

$$\frac{\partial T_\Omega(\Gamma_r, \tilde{\omega})}{\partial t} = \beta \int d\omega\, d\omega'\, D_R(\omega, \omega') \langle h(\omega') \rangle \hat{f}(\tilde{\omega} - \omega') + \xi'(t, \tilde{\omega}) \tag{A.8}$$

where

$$\xi'(t, \tilde{\omega}) = \int_\Omega d^3R\, d\omega\, \xi(t, \bar{R}, \omega) \hat{f}(\tilde{\omega} - \omega), \tag{A.9}$$

and the rotational diffusion coefficient $D_R$ is given by

$$D_R(\omega, \omega') = \int_\Omega d^3R\, d^3R'\, D(\bar{R}, \bar{R}', \omega, \omega'). \tag{A.10}$$

We suggest that the Langevin equation described here is more appropriate. For example, a Smoluchowski equation for $T_\Omega$ as presented in [37-38] implies a Langevin equation for the evolution of the set of polar angles. However, the latter does not qualify as an order parameter because its dynamics is on a similar timescale to that of atomistic fluctuations which violates the arguments needed to derive such an equation. Furthermore, our approach is calibration-free and can be readily generalized to more complex cases supporting a rich array of liquid-crystal phenomena.

**Acknowledgements**

This project was supported in part by the National Institute of Health (NIBIB), the National Science Foundation (CRC Program), and Indiana University's college of Arts and Sciences through the Center for Cell and Virus Theory.